\documentclass[a4paper,12pt]{article}
\begin{document}
\pagestyle{plain}
\title{Parametrization of the quark mixing matrix involving its eigenvalues}
\author{ S. Chaturvedi \thanks{scsp@uohyd.ernet.in}\\
\rm School of Physics, University of Hyderabad, \\
 Hyderabad 500 046 India\\
Virendra Gupta \thanks{virendra@aruna.mda.cinvestav.mx}\\
\rm Departamento de Fis\'ica Aplicada, CINVESTAV-Unidad M\'erida\\
 A.P. 73 Cordemex 97310 M\'erida, Yucatan, Mexico}
\date{\today}
\vskip-1cm
\maketitle
\begin{abstract}
A parametrization of the $3\times 3$ Cabibbo-Kobayashi-Maskawa  matrix, 
$V$, is presented in which the parameters are the eigenvalues and the 
components of its eigenvectors. In this parametrization, the small 
departure of the experimentally determined $V$ from being moduli symmetric 
(i.e. $|V_{ij}|=|V_{ji}|$) is controlled by the small difference between 
two of the eigenvalues. In case, any two eigenvalues are equal, one obtains 
a moduli symmetric $V$ depending on only three parameters. Our parametrization 
gives very good fits to the available data including CP-violation. Our value 
of $\sin 2\beta\approx 0.7$ and other parameters associated with the `
unitarity triangle' $V_{11}V_{13}^{*}+V_{21}V_{23}^{*}V_{31}V_{33}^{*}=0$ are
in good agreement with data and other analyses.
\end{abstract}
\newpage
\section{Introduction}
Flavor mixing of the quarks, in the Standard Model, is understood through 
the Cabibbo-Kobayashi-Maskawa (CKM) matrix. Since the first \cite{1} explicit
parametrization of the CKM matrix, $V$, for three generations, many different 
parametrizations have been suggested \cite{2,3}. In all these cases, the
mixing matrix $V$ is parametrized in terms of four parameters, three angles and
a phase. However, other approaches to the parametrization are possible and
available \cite{4,5,6}. 

Any $3\times 3$ unitary matrix $V$ can be expressed in terms of its
eigenvalues $E_i=\exp(i\alpha_i)~,~i=1,2,3$ as follows, 
\begin{equation}
V=W{\hat V}W^\dagger,
\label{1}
\end{equation}
where ${\hat V}={\rm diag}(E_1, E_2, E_3)$ and the
diagonalizing matrix $W$ is unitary. In reference [4], arguments were
presented that due to re-phasing freedom, the eigenvalues can be chosen 
at will and
there they were fixed to be the three roots of unity, so that $V$ depended on the
four parameters needed to specify $W$. For confrontation of data, $W$ was
chosen to depend on only two parameters which resulted in a symmetric $V$
($ {\rm i. e.}, V_{ij} = V_{ji}$) and which gave a reasonable fit to the data
available at that time.

In the approach of reference \cite{6}, the CKM matrix was parametrized as 
\begin{equation}
V(\theta) = \cos\theta ~I + i \sin\theta~ U .
\label{2}
\end{equation}
The parameter $\theta$ determines the relative importance of the trivial 
part, $I$ {\it vis a vis} the non-trivial part $U$. The hermitian and unitary 
$U$ (independent of $\theta$) depends on two real positive parameters 
\cite{7}. Since $U=U^\dagger$, $V$ is moduli symmetric (i.e.$|V_{ij}|=
|V_{ji}|$). Such a matrix can always be made symmetric by rephasing 
\cite{7,8} and in general has only three parameters. Such a $V$ gave a good 
fit to the available data \cite{3} though its predictions for $\rho, \eta$ 
and $\sin2\beta$ were on the larger side compared to the recently available 
data \cite{3}. 

In this paper we consider a parametrization of $V$ based on Eq$(\ref{1})$ for
general eigenvalues, (that is as explicit parameters), even though it is clear
that they have no physical significance.  As we shall see, such a
parametrization exhibits different features of the mixing matrix not
accessible otherwise. In particular, we also are motivated to have the 
eigenvalues 
as explicit parameters because there may be an underlying connection between 
the eigenvalues of the mixing matrix in the quark and lepton sectors. The 
first hint of this came from the application of the approach of Eq$(\ref{2})$ 
to the neutrino mixing matrix $V_{\nu}$\cite{10}. Writing 
\begin{equation}
V_\nu(\theta_\nu) = \cos\theta_\nu~I + i \sin\theta_\nu ~U_\nu,
\label{3}
\end{equation}
one finds that the maximal mixing of $\nu_\mu$ and $\nu_\tau$ (indicated by 
the atmospheric neutrino data \cite{10}) requires $\theta_\nu=\pi/4$. The fit
to the CKM data gave $\theta=\pi/4$!. The remarkable equality 
$\theta=\theta_\nu=\pi/4$ suggests an underlying quark-lepton symmetry in this
approach, even though the full $V$ and $V_\nu$ are very different. The really
interesting point is to realize that in these parametrizations the parameters 
$\theta$ and $\theta_\nu$ completely determine the eigenvalues of $V$ and
$V_\nu$ respectively! In fact, the actual eigenvalues of $V$ in 
Eq$(\ref{2})$ are $exp(i\theta), exp(-i\theta)$ and $exp(-i\theta)$ 
while the two real parameters in $U$ determine the corresponding
eigenvectors. This applies {\it mutatis mutandis} to $V_\nu$. In general, if
two eigenvalues are equal it follows that $V$ can depend on at most three
parameters and can be made symmetric (see Section II below and \cite{8}). 

By considering the general case, when the three eigenvalues are different, we
can obtain a $V$ which is `asymmetric' (i.e. $|V_{ij}|\neq |V_{ji}|$). 
Experimentally, this asymmetry is quite small and in fact the major 
part of $V$ is indeed given by the parametrization of the form 
Eq$(\ref{2})$. In the `eigenvalue parametrization' which we consider the 
small asymmetry is contributed by the small difference between two 
eigenvalues. By 
confronting this and other possible ways of parametrizing the CKM matrix one 
can hope to obtain a better understanding of the nature and structure of the 
quark mixings. 

In Section II, Eq$(\ref{1})$  is considered in detail and the general notation,
formulae and their consequences are given. In Section III  we consider the
confrontation of the eigenvalue parametrization with data for simplified
choices of $W$. Numerical results for the fits are presented in Section IV. 
Finally we conclude with a brief summary and remarks in Section V.

\section{General eigenvalue parametrization of V}

Our starting point is Eq$(\ref{1})$. We can write it as
\begin{equation}
V=\sum_{k=1}^{3} E_k~N_k,
\label{4}
\end{equation}
where $N_k$ are the `projectors' for $V$. They satisfy $\sum_{k=1}^{3}N_k=I~,~
N_k= N_{k}^{\dagger}~,~N_kN_{k^\prime}=N_k\delta_{kk^\prime}$ and 
$(N_k)_{lm}= W_{mk}^{*}W_{lk}~,~l,k=1,2,3$ where $W_{lk}$ are matrix elements 
of the matrix $W$. We can choose the overall phase of $V$, in general, so 
that on eliminating $N_3$, we have 
\begin{equation}
V= I+F_1 N_1 +F_2 N_2,
\label{5}
\end{equation}
where
\begin{equation}
F_i= (E_i-1)=(\exp(i\alpha_i)-1), i=1,2 .
\label{6}
\end{equation}
The choice $\alpha_3=0$ or $E_3=1$ has been made in Eqs$(\ref{5},\ref{6})$. 
The columns 1 to 3 of $W$ are the orthonormal eigenvectors of $V$ for 
eigenvalues $E_1$ to $E_3$. Consequently we can write the hermitian 
projection matrices as 
\begin{equation}                          
N_k= \left(\,
\matrix{
c_k \cr
b_k\cr
a_k}\,\right) \otimes
\left(\matrix{c_{k}^{*},& b_{k}^{*},& a_{k}^{*}}\right)
=\left(\,
\matrix{ |c_k|^2&c_kb_{k}^{*}&c_{k}a_{k}* \cr
b_{k}c_{k}^{*}&|b_k|^2&b_{k}a_{k}^{*} \cr
a_kc_{k}^{*}&a_kb_{k}^{*}&|a_k|^2}\,\right),  
\label{7}
\end{equation}
where we have introduced the compact notation $(W_{1k},W_{2k},W_{3k})= 
(c_k,b_k,a_k)$ for $k=1,2,3$. Our parametrization is based on 
Eqs$(\ref{5}-\ref{7})$. It is the generalization of Eq$(\ref{2})$ to which 
Eq$(\ref{5})$ reduces when two eigenvalues of $V$ are equal. 

In Eq$(\ref{5})$, $V$ depends on two eigenvalues $E_1$ and $E_2$ or 
equivalently on the real parameters $\alpha_1$ and $\alpha_2$. The matrices 
$N_1$ and $N_2$ seemingly depend on six complex numbers $a_k,b_k,c_k~,~k=1,
2$. However, by simple rephasing we can make $N_1$ real, that is, take 
$a_1,b_1$ and $c_1$ 
to be real and positive. Further, we can choose one (non-zero) component of
the eigenvector for $E_2$ to be real. We choose $c_2$ to be real. The
unitarity of $W$ (or the orthonormality of the eigenvectors) gives us 
\begin{equation}
a_1a_{2}^{*}+b_1b_{2}^{*}+c_1c_{2}^{*} =0, \label{8a}
\end{equation}
\begin{equation}
|a_1|^2+|b_1|^2+|c_1|^2=1,
\end{equation}
\begin{equation}
|a_2|^2+|b_2|^2+|c_2|^2=1.
\end{equation}
These equations show that each $N_1$ and $N_2$ depends on two real parameters
or $W$ depends on four real parameters. These can, for example, be taken to be 
$|a_1|,|b_1|$and   $|a_2|,|b_2|$. Note that Eq$(\ref{8a})$ will determine the 
real and imaginary parts of $a_2$ and $b_2$ ($c_1, c_2$ being real). Thus, 
$V$ in Eq$(\ref{5})$ depends seemingly on six parameters, $\alpha_1$ and 
$\alpha_2$ which determine the eigenvalues plus the four in $N_1$ and $N_2$
which determine the corresponding eigenvectors. 

Before confronting data, we discuss some general consequences of
Eq$(\ref{5})$ pertaining to the asymmetry of $V$. For a unitary matrix the
departure from moduli symmetry (i.e. $|V_{ij}|=|V_{ji}|$) is conveniently given
by the formula
\begin{eqnarray}
\Delta(V)&\equiv& |V_{12}|^2-|V_{21}|^2=|V_{23}|^2-|V_{32}|^2
=V_{31}|^2-|V_{13}|^2 \nonumber\\
&=& -16 \sin(\frac{\alpha_1-\alpha_2}{2})\sin(\frac{\alpha_1-\alpha_3}{2})
\sin(\frac{\alpha_2-\alpha_3}{2})J(W),
\label{9}
\end{eqnarray}
where
\begin{equation}
J(W)= {\rm Im }(W_{11}W_{21}^{*}W_{12}^{*}W_{22})= 
{\rm Im}(c_1b_{1}^{*}c_{2}^{*}b_2),
\label{10}
\end{equation}
is the Jarlskog \cite{11} invariant for the matrix $W$ in our notation. 

This formula ( see also \cite{8}) determines the conditions when $V$ is moduli
symmetric.

\noindent
(i) If $J(W)=0$ then $|V_{ij}|=|V_{ji}|$ even though all the eigenvalues are
different. In our choice of parameters above, since $a_1, b_1, c_1$ and $c_2$
are real, it is imperative that $a_2$ and $b_2$ have an imaginary part so that 
$J(W)\neq 0$ and Eq$(\ref{5})$ would give an asymmetric $V$.

\noindent
(ii) If any two eigenvalues of $V$ are equal then $|V_{ij}|=|V_{ji}|$ even
though $J(W)\neq 0$.  If we take $E_2=E_3=1$ i.e. $\alpha_2=\alpha_3=0$ and put
$\alpha_1=2\theta$ then with a little manipulation Eq$(\ref{2})$ can be
written as $V(\theta)\exp(i\theta)= I+(F_1-1)N_1$ where the parameters $a,b,c$ 
of reference\cite{6} are written as $
a=-i|a_1|\exp[i(\phi_{b_{1}}-\phi_{c_{1}})], 
b= i|b_1|\exp[i(\phi_{a_{1}}-\phi_{c_{1}})], 
c=-i|c_1|\exp[i(\phi_{a_{1}}-\phi_{b_{1}})]$. 

The main point is that whenever two eigenvalues are equal we can write the
unitary matrix $V$ as $ V=I+\lambda N$ where
$\lambda=2i\sin\theta\exp(i\theta)$ and $N=N^{\dagger}=N^2$ \cite{12}. 

It is very interesting that from the eigenvalue parametrization approach, the 
parametrization of Eq$(\ref{2},\ref{3})$ turns out to be a particular case of 
Eq$(\ref{5})$ even though the original motivation for Eq$(\ref{2},\ref{3})$ 
was quite different. Furthermore, the interesting parameters $\theta$ and 
$\theta_\nu$ turn out to be eigenvalues of the two mixing matrices $V$ and 
$V_{\nu}$ respectively.

In the next section we go on to confront Eq$(\ref{5}-\ref{7})$ with available
data which shows that there is a small asymmetry in $V$, that is 
$\Delta(V)\neq 0$ though small.

\section{Numerical Results}
Experiments can only determine $|V_{ij}|$ for us. Since $V$ is unitary, four
independent moduli are sufficient to determine all the nine $|V_{ij}|$. This
implies that we can have many different parametrizations of the complete
complex matrix $V$ as long as they give the same $|V_{ij}|$ in agreement with
experiments. 

The Particle Data Group gives experimentally determined ranges for
$|V_{ij}|$. One can convert these ranges into a central value with errors and
use these for fitting. This procedure has a draw back that the unitarity
constraints in the moduli are not exact. Instead we use the `standard' 
parametrization (Eq$(11.3)$, Section 11 in \cite{3}) to fit the
moduli. Accordingly, we take \cite{3}, $s_{12}=0.2229\pm 0.0022$, $s_{23}=
0.0412\pm 0.0020$ and $s_{13}=0.0036 \pm 0.0007$ with $\delta_{13}
=59^\circ \pm 13^\circ= (1.02\pm 0.22){\rm radians}$. This gives the moduli 
matrix $V_{mod}=(|V_{ij}|)$ to be 
\begin{equation}
V_{mod}
=\left(\,
\matrix{ 0.974835\pm 0.000503&0.222899\pm 0.002199&0.0036\pm 0.0007 \cr
0.222786\pm 0.002198&0.973996\pm 0.000509&0.0411997\pm 0.001999 \cr
0.00793254\pm 0.000877&0.0405888\pm 0.0019569&0.999144\pm 0.0000825}\,\right).
\label{11}
\end{equation}
We confront the central values to determine the parameters in our 
parametrization. The advantage of using Eq$(\ref{11})$ is that the 
unitarity constraints on $|V_{ij}|$ are satisfied.

As noted earlier, our parametrization in Eq$(\ref{5})$ has six real parameters
while the data gives only four independent inputs, namely four $|V_{ij}|$'s. 
There are two obvious ways to reduce the parameters. 

\noindent
[A] Keep the two eigenvalues of $V$ (i.e. $\alpha_{1}$ and $\alpha_2$) as
parameters and choose a $W$ with only two parameters. That is, the
corresponding eigenvectors are determined by only two real numbers. We explore
this possibility here. We will see that a small $\alpha_2$ (implying the
eigenvalue $E_2 \approx 1-i\alpha_2$ is near $E_3=1$) controls the asymmetry 
in $V$. 

\noindent
[B] The other way is to arbitrarily choose the eigenvalues in advance (e.g. in 
\cite{4}) and then determine the parameters which determine the eigenvectors
of $V$. We will consider this briefly for sake of comparison.
\vskip0.7cm
\noindent
{\bf Type A fits}

In these fits the $W$ used has only two parameters which determine the
eigenvectors of $V$. To obtain such a $W$ we consider a general four parameter
$W$ and reduce the parameters to two guided by notions of simplicity. 
\vskip0.5cm
\noindent
{\it Case(i)}
We start with a $W$ parametrized by three angles $\theta_{12}, \theta_{13}, 
\theta_{23}$ and a phase $\delta_{13}$ as in Eq$(11.3)$, Section 11 of 
\cite{3}. If one requires that $W$ be moduli symmetric i.e. $|W_{ij}|= 
|W_{ji}|$ then one needs two conditions, namely $\theta_{13}=\theta_{23}$ 
and $-2\cos \delta_{13}=\cot\theta_{12}\tan\theta_{23}\sin\theta_{23}$. 
This results in a two parameter $W$ which we refer to as $W_P$. Using $W_P$ 
and the two parameters 
$\alpha_1$ and $\alpha_2$ from the eigenvalues we make a four parameter 
fit to the $|V_{ij}|$ given in $V_{mod}$ in Eq$(\ref{11})$. One obtains an
excellent fit, the numerical values of the parameters are given in Table I. 
Note that in this case only $c_1$ and $c_2$ are real but $a_i, b_i,~i=1,2$ 
are complex. Consequently both the matrices $N_1$ and $N_2$ are complex 
though given in terms of two real parameters.  
\vskip0.5cm
\noindent
{\it Case(ii)}
In this case, we consider the parametrization of $W$ a la Kobayashi-Maskawa 
( \cite{1} or Eq$(11.4)$ in Section 11 of \cite{3}). Here $W$ becomes moduli
symmetric if one simply takes $\theta_2=\theta_3$, so to reduce the
parameters to two we make the choice, $\delta=\pi-\theta_1$. This gives 
$\cos\delta=-\cos\theta_1$ and $\sin\delta=\sin\theta_1$. This two parameter
W, denoted by $W_{KM}$ is quite simple
\begin{equation}
W_{KM}
=\left(\,
\matrix{ C_1&-S_1C_2&-S_1S_2 \cr
S_1C_2&C_1-iS_{2}^2S_1&iS_1C_2S_2 \cr
S_1S_2&iS_1C_2S_2&C_1-iC_{2}^2S_1}\,\right),  
\label{12}
\end{equation}    
where $C_i\equiv \cos\theta_i,~S_i\equiv \sin\theta_i,~i=1,2$.
This case, has the feature that ${\rm Re}~a_2=0$. The numerical values of the
parameters $a_1=S_1S_2, a_2=iS_1C_2S_2$ etc are given in Table I. Note that 
except for $a_2$ and $b_2$ others are real, so that $N_1$ is real but 
$N_2$ is not. Again the fit to $|V_{ij}|$ is excellent and the value of the 
parameters in the two fits differ only slightly. 

The main contribution to the CP-violation parameter 
$J(V)={\rm Im}(V_{11}V_{22}V_{12}^{*}V_{21}^{*})$ comes from the first two
terms of Eq$(\ref{5})$ which give a moduli symmetric $V$ (the limit $\alpha_2 
~{\rm or}~ F_2 \rightarrow 0)$. In fact, for the complete $V$, $
J(V)=2.744\times 10^{-5} $ while it becomes equal to $2.87\times10^{-5}$ 
when $\alpha_2=0$. The small value of $\alpha_2= -0.106365 ~{\rm radians}$ 
(compared to $\alpha_1= 1.88053~ {\rm radians}$) gives the small asymmetry, 
$\Delta(V)= 5\times 10^{-5}$. 
\vskip0.7cm
\noindent
{\bf Type B fits}

In these fits one specifies or chooses the eigenvalue parameters $\alpha_1$
and $\alpha_2$ and then determines the four parameters in $W$ from the
data. These can be chosen to be $|a_i|, |b_i|,i=1,2$. These determine the
imaginary parts of $a_2$ and $b_2$ through Eq$(\ref{8a})$ since $c_2$ is
real. For comparison with above results, we choose $\alpha_1=-\alpha_2=
120^\circ$ following reference\cite{4}. The choice of $\alpha_1$ and
$\alpha_2$ can not be completely arbitrary because of the numerical values of 
$|V_{ij}|$ determined experimentally. For example, the choice\cite{4} 
${\rm Tr} V=0$ implies inequalities like $ ||V_{22}|-|V_{33}||\leq 
|V_{11}|\leq|V_{22}|+|V_{33}|$ etc.. These are satisfied by the data but will
not be valid for every unitary matrix. The numerical results are given in
Table I. The fit is as good as Type A fit but there is no obvious reason why 
$\Delta(V)$ is so small because the contribution of $F_1N_1$ and $F_2N_2$ are
of the same order. Note that $|F_1|=|F_2|$ and $|(N_1)_{ij}|\approx 
|(N_2)_{ij}|$. The phases of $(N_k)_{ij}$ conspire in some way to give a small asymmetry.

In conclusion, we consider the Type A fits to be more meaningful as they
display the structure of $V$, that it is mainly moduli symmetric with a small 
parameter monitoring the small asymmetry. As can be seen from Table I the 
parameters $a_i,~b_i,~c_i$ for the type A case are very similar, particularly 
their moduli. In fact, $|a_2|\approx |a_1|,~|b_2|\approx |c_1|$ and 
$|c_2|\approx |b_1|$, so that, the matrix elements of $N_1$ and $N_2$ are 
comparable. However,  their coefficients $F_1=(-1.3048+0.952415~i)$ 
and $F_2=(-0.00565138-0.106164~i)$ are very different. Since $|F_1|\approx
1.6$ and $|F_2|\approx 0.1$, the $I+F_1N_1$ part gives the major contribution 
to $V$ while the much smaller $F_2N_2$ contributes to give a small asymmetry.
  
\section{Predictions for the parameters of the unitarity triangle}
 The unitarity constraint 
\begin{equation}
V_{11}V_{13}^{*}+V_{21}V_{23}^{*}+V_{31}V_{33}^{*}=0,
\label{13}
\end{equation}
can be written as $z_1+z_2+z_3=0$ where $z_i=V_{i1}V_{i3}^{*}, i=1,2,3$. The
angles of this triangle, in standard notation, are 
$\alpha={\rm arg}(-z_3/z_1),\beta={\rm arg}(-z_2/z_3)$ and 
$\gamma={\rm arg}(-z_1/z_2) $. These can be determined directly from our fits
and are given in Table II. In addition, the values of $\rho$ and $\eta$ 
(defined as $-z_1/z_2=\rho+i\eta$) are also given. They are connected to the
angles through
\begin {equation}
  \sin\alpha=\frac{\sin\beta}{\sqrt{\rho^2+\eta^2}}=
   \frac{\sin\gamma}{\sqrt{(1-\rho)^2+\eta^2}};~~
         \tan\gamma = \eta/\rho .
\label{14}
\end{equation}
The $\rho$ and $\eta$ defined here and Eq$(\ref{14})$ are valid for any exact 
parametrization of the CKM matrix.

Since we fit $|V_{ij}|$ given in Eq$(\ref{11})$, that is, the inputs for all
the three fits are the same, we obtain the same values for the angles, $\rho$ 
and $\eta$ in all the cases, These are given in column 1, Table II. These
values are to be compared to those obtained by the Particle Data Group 
\cite{3}, namely, $\beta=24^\circ\pm 4^\circ$, $\gamma=59^\circ\pm 13^\circ$ 
and ${\bar \rho}=0.22\pm 0.10$ and ${\bar \eta}=0.35\pm 0.05$. There is a very
minute numerical difference between $(\rho, \eta)$ and $({\bar \rho}, 
{\bar \eta})$. For the definition of the latter see reference \cite{13}. The
agreement between their and our values is quite satisfactory. For comparison, 
the 
second column of Table II, gives the values obtained when $\alpha_2=0$, that 
is, when two eigenvalues are equal, $E_2=E_3=1$ and $V$ is symmetric. As one 
can see the values of $\eta$ and particularly $\rho$ are higher. Also, there 
is a change in the values of the angles by about $10-20^\circ$. This change 
can be seen more clearly in the values of $\sin 2\beta$ in the two cases. For 
the symmetric case, $\sin 2\beta=0.9532$ compared to $0.6988\approx 0.7$ for 
the asymmetric (or actual) $V$. These are to be compared to the measured 
\cite{14} value $\sin 2\beta = 0.78\pm 0.08$. Our value $0.7$ is in reasonable
agreement. Experiments are in progress both at Belle and BaBar for a better 
value of $\sin 2\beta$ and to measure $\sin 2\alpha$. We should have a 
clearer picture in a couple of years. 

\section{Summary and final remarks}

In the parametrization considered here the parameters directly determine the
mathematical structure of the CKM matrix $V$, namely its eigenvalues $E_i$ and 
its eigenvectors. Such a parametrization brings out the fact that
experimentally $V$ is practically  moduli symmetric, the small asymmetry
observed is due to two eigenvalues being very close to each other. It is
intriguing to note that smallness of the asymmetry measured by $\Delta(V)=
5.00\times 10^{-5}$ is of the same order as $J(V)= 2.74\times 10^{-5}$, even 
though, in general, they are not related. 

Finally, extension of this approach
to the lepton sector in the near future would be of much interest. Furthermore 
our approach can be easily extended for the case of four or more generations. 

\vskip2.0cm 
\noindent
{\bf Acknowledgments}
One of us (VG) would like to thank CONACyT, Mexico for its support under 
Project no. 32598PE. He would also like to thank the University of Hyderabad, 
India for the award of the Jawaharlal Nehru Chair during July-September, 
2002 and the School of Physics, University of Hyderabad for its hospitality 
during his stay.

\newpage
\begin{tabular}{|c|c|c|c|}\hline
 & Type A Case (i) &Type A Case(ii)&  Type B \cr\hline
$\alpha_1$ & $107.748^\circ$&$107.746^\circ$&$120^\circ$  \\ \hline
$\alpha_2$ &$-6.09428^\circ$&$-6.09424^\circ$&-$120^\circ$  \\ \hline
$c_1$&$0.134232$&$0.134234$&$0.129796$  \\ \hline
$b_1$&$-0.99062-0.0000876785~i$&$0.99062$&$0.99126$  \\\hline
$a_1$&$0.025322-0.00343008~i$&$0.0255536$&$-0.0235971$  \\ \hline
$c_2$&$0.99062$&$-0.99062$&$0.991539$  \\\hline
$b_2$&$0.134232-0.000647057~i$&$0.134234-0.00065895~i$&
$-0.129753+0.0000376876~i$  \\ \hline
$a_2$&$-0.00343008-0.0253136~i$&$0.0255451~i$&$0.00333381+0.00158317~i$  \\ \hline
\end{tabular}
\vskip0.5cm
Table I : Parameters in $V$ (Eq$(\ref{5})$) determined by fitting the central 
values of $|V_{12}|,~|V_{21}|,~|V_{13}||V_{23}|~$ in Eq$(\ref{11})$ as 
inputs. Note that the eigenvalues $E_1$ and 
$E_2$ ( or $\alpha_1$ and $\alpha_2$) and the moduli $|a_i|,~|b_i|$ and 
$|c_i|$ for the Type A fits are practically the same. We have kept the 
number of places of decimal as given by the mathematica program. This ensures
that unitarity etc. constraints are obeyed exactly and also this shows 
where slight difference in the values of parameters occurs in the different 
fits. See text for the definition of the parameters and details.
\vskip1.5cm
\begin{tabular}{|c|c|c|}\hline
        &Type A &  Type A , $\alpha_2=0$ \cr\hline
$ \rho$ & $0.200284$&$0.4911$  \\ \hline
$\eta$  & $0.325685$&$0.3719$  \\ \hline
$\alpha$&$99.43^\circ$&$106.70^\circ$  \\ \hline
$\beta$ &$22.16^\circ$&$36.16^\circ$  \\\hline
$\gamma$&$58.41^\circ$&$37.13^\circ$  \\ \hline
$\sin2\beta$&$0.6986$&$0.953 $  \\ \hline
$\sin2\alpha$&$-0.3232 $&$-0.550 $  \\ \hline
\end{tabular}
\vskip0.5cm
Table II: Values of the angles $(\alpha,~\beta,~\gamma)$ and parameters 
$\rho$ and $\eta$ connected with the unitary triangle (Eq$(\ref{13})$) are 
given. For the Type A fits both Cases (i) and (ii) give practically the same 
values. Column 1 gives the values for the asymmetric $V$ ($\alpha_2\approx 
-6.1^\circ$) while column 2 gives the values for the moduli symmetric 
case ($\alpha_2=0^\circ$).  


\begin{thebibliography}{99}
\bibitem{12} This is obvious from the `Projector Representation' (PR) 
in Eq$(\ref{4})$ or Eq$(\ref{5})$. In fact the PR for the $n\times n$ case 
tells us that $V$ is moduli symmetric and depends on $n$ real parameters 
if it has only two independent eigenvalues. For `asymmetry' it should have at 
least three or more different eigenvalues. For example, for $n=4$, $V$ is 
moduli symmetric for the cases $E_1\neq E_2=E_3=E_4$ amd $E_1=E_2\neq E_3=E_4$ 
but can be asymmetric $E_1\neq E_2\neq E_3=E_4$ or if all four eigenvalues 
are different.
\end{thebibliography}
\end{document}